\documentclass[conference]{IEEEtran}
\IEEEoverridecommandlockouts
\usepackage{cite}
\usepackage{amsmath,amssymb,amsfonts}
\usepackage{graphicx}
\usepackage{textcomp}
\usepackage{xcolor}
\def\BibTeX{{\rm B\kern-.05em{\sc i\kern-.025em b}\kern-.08em
    T\kern-.1667em\lower.7ex\hbox{E}\kern-.125emX}}
\usepackage{oswald_custom}
\usepackage{spverbatim}
\usepackage{listings}
\usepackage{siunitx}
\usepackage{algpseudocode}
\usepackage{algorithm}

\usepackage{multirow}
\usepackage{hyperref}
\definecolor{darkblue}{rgb}{0, 0, 0.5}
\hypersetup{
    colorlinks=true,
    citecolor=darkblue,
    linkcolor=darkblue,
    urlcolor=darkblue
}
\usepackage{cprotect}
\usepackage{booktabs}

\makeatletter
\newcommand{\newlineauthors}{%
  \end{@IEEEauthorhalign}\hfill\mbox{}\par
  \mbox{}\hfill\begin{@IEEEauthorhalign}
}
\makeatother

\begin{document}

\title{Predictive Speech Recognition and End-of-Utterance Detection Towards Spoken Dialog Systems}

\author
{
\IEEEauthorblockN{
    Oswald Zink$^{\star,\dagger}$\thanks{$^*$This work was conducted while the first author was at Waseda University.},
    Yosuke Higuchi$^{\star}$,
    Carlos Mullov$^{\dagger}$,
    Alexander Waibel$^{\dagger}$ and
    Tetsunori Kobayashi$^{\star}$
}
\IEEEauthorblockA{
    \textit{$^{\star}$Waseda University, Tokyo, Japan},
    \textit{$^{\dagger}$Karlsruhe Institute for Technology, Karlsruhe, Germany}
}
}

\maketitle

\begin{abstract}
Effective spoken dialog systems should facilitate natural interactions with quick and rhythmic timing,
mirroring human communication patterns.
To reduce response times,
previous efforts have focused on minimizing the latency in automatic speech recognition (ASR) to optimize system efficiency.
However, this approach requires waiting for ASR to complete processing until a speaker has finished speaking,
which limits the time available for natural language processing (NLP) to formulate accurate responses.
As humans, we continuously anticipate and prepare responses even while the other party is still speaking.
This allows us to respond appropriately without missing the optimal time to speak.
In this work,
as a pioneering study toward a conversational system that simulates such human anticipatory behavior,
we aim to realize a function that can predict the forthcoming words and estimate the time remaining until the end of an utterance (EOU),
using the middle portion of an utterance.
To achieve this, we propose a training strategy for an encoder-decoder-based ASR system,
which involves masking future segments of an utterance and prompting the decoder to predict the words in the masked audio.
Additionally, we develop a cross-attention-based algorithm
that incorporates both acoustic and linguistic information to accurately detect the EOU.
The experimental results demonstrate the proposed model's ability to predict upcoming words and estimate future EOU events up to 300ms prior to the actual EOU.
Moreover, the proposed training strategy exhibits general improvements in ASR performance.

\end{abstract}
\section{Introduction}
\label{sec:intro}
In natural human conversations,
individuals frequently begin speaking immediately after or even before the other party finishes.
This rhythmic and prompt flow of dialog is facilitated by their speculative ability to predict what the other will say next and anticipate when they will stop speaking,
thus allowing for the formulation of timely and appropriate responses.
In contrast, current spoken dialog systems typically lack this capability,
as they only begin to prepare responses after their automatic speech recognition (ASR) module has completed its transcriptions.
This delay forces subsequent NLP modules --- language understanding, dialog policy, and natural language generation --- to produce responses within a limited timeframe,
potentially compromising the quality and speed of interactions with users.

To address the delay inherent in ASR,
recent research has focused on minimizing the latency of ASR systems,
developing online streaming architectures~\cite{graves2012sequence,he2019streaming,zhang2020transformer,moritz2020streaming,tsunoo2021streaming,zhao2023mask}
that effectively control the number of look-ahead frames~\cite{narayanan2021cascaded,li2021better,moritz2021dual,zhao2023conversation,yu2021dual,strimel2023lookahead}.
Nonetheless, these models typically operate in real-time,
requiring input audio information up to the final end of an utterance (EOU) frame.
As a result, the overall latency of a dialog system experienced by users is inevitably above zero seconds.
Simultaneously, the downstream NLP modules face challenges in achieving quick responses within the desired latency window,
which can span from as early as \SI{-200}{\milli\second} to \SI{400}{\milli\second} in natural human interactions~\cite{levinson2015timing,ekstedt2020turngpt,sakuma2023response}.

Inspired by the speculative capabilities of humans,
there have been efforts to anticipate future information in incomplete utterances,
extending beyond the constraints of conventional real-time ASR systems.
A predominant approach involves using an external language model to generate forthcoming words from a partial ASR hypothesis,
as discussed in studies~\cite{ekstedt2020turngpt,sakuma2023response,yusuf2024speculative}.
In~\cite{ekstedt2020turngpt,sakuma2023response},
the syntactic completeness of spoken text at mid-utterance is evaluated to represent the likelihood that the speaker will continue speaking,
which has proven crucial for efficiently predicting turn-shifts and timing responses.
Similarly, \cite{yusuf2024speculative} has fine-tuned a large language model to extrapolate words from incomplete ASR outputs,
effectively incorporating audio information from an ASR model through a soft prompting technique.

This work builds upon these foundational efforts to advance ASR systems with predictive functionality,
aiming to provide the downstream NLP modules with ample time to operate and enable the dialog system to respond effectively.
Uniquely, our approach functions as a straightforward extension of traditional ASR models,
which can naturally incorporate both acoustic and linguistic contexts to enable prediction.
Additionally, we avoid reliance on external language models, particularly the recent large language models,
which can impose significant computational demands for the front end of dialog systems.
To this end, we propose an encoder-decoder-based ASR model~\cite{chorowski2015attention,chan2016listen,watanabe2017hybrid}
that is specifically constructed to simulate human predictive capabilities within a unified framework,
designing predictive EOU detection and predictive ASR tasks.
Predictive EOU detection forecasts the future endpoint of an utterance.
Predictive ASR, on the other hand, generates the complete transcription before the utterance concludes,
thereby allowing the downstream NLP modules to start processing earlier for a low-latency response~\cite{chang2020low}.
More concretely, for predictive EOU prediction,
our approach leverages alignment information obtained from the cross-attention mechanism.
We feed mid-utterance input followed by empty input (only positional embeddings) into the model, and
by analyzing the attention weights applied to the empty input, we predict future endpoints.
For predictive ASR, we use the decoder to produce tokens corresponding to the empty input,
where the decoder functions as a generative language model to predict future tokens.
In order for the model to operate even in the absence of future speech input,
we design a training strategy that randomly masks segments of future speech,
thereby facilitating the training of the language model capability in the decoder.

The remainder of this paper is organized as follows.
Section~\ref{sec:methods} details the proposed ASR model,
which aims to enhance the predictive abilities for efficient dialog systems.
Section~\ref{sec:experiments} evaluates the efficacy of our model through speech recognition experiments,
focusing on the trade-offs between performance and latency.
Finally, Section~\ref{sec:conclusion} concludes this paper.

\section{Predictive EOU Detection and Predictive ASR}
\label{sec:methods}
This section presents the proposed ASR model,
designed to support predictive EOU detection and predictive ASR capabilities.
These features enable the model to complete its process before an utterance concludes,
thus ensuring the NLP modules within dialog systems have adequate preparation time to generate responses.
The subsequent sections first detail the two predictive tasks of interest (as outlined in Section~\ref{sec:intro}),
followed by an explanation of how these capabilities are trained and applied for use.

\begin{figure}
    \centering
    \includegraphics[width=0.85\linewidth]{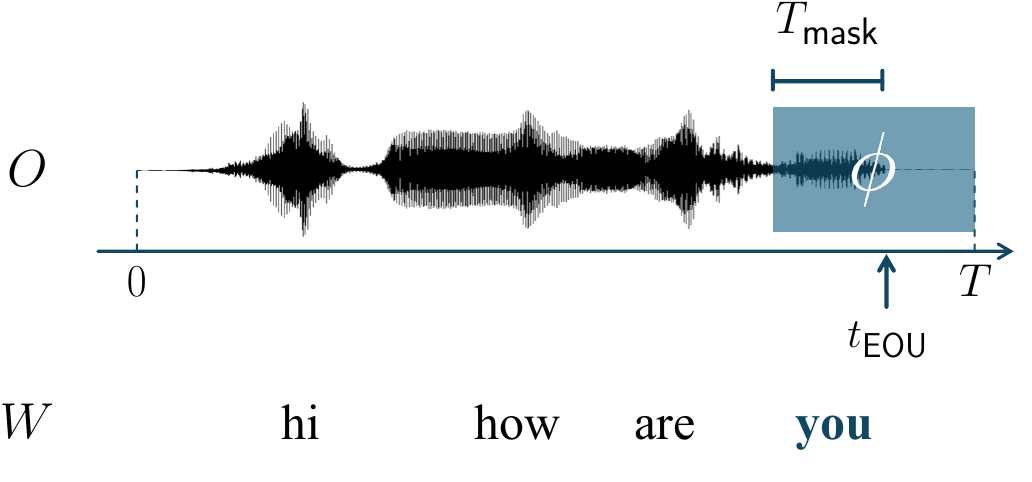}
    \vspace{-0.3cm}
    \caption{Schematic drawing of predictive tasks of interest. Given an utterance of which we mask $T_{\mathsf{mask}}$ milliseconds ahead of EOU and the trailing silence shown as $\phi$, predictive EOU detection tries to predict $t_{\mathsf{EOU}}$ based on the available audio information. The goal of predictive ASR is to generate words corresponding to the masked input (i.e., ``you'') based on the visible audio information and preceding tokens.}
    \label{fig:task}
    \vspace{-0.1cm}
\end{figure}

\begin{figure}
    \centering
    \includegraphics[width=0.98\linewidth]{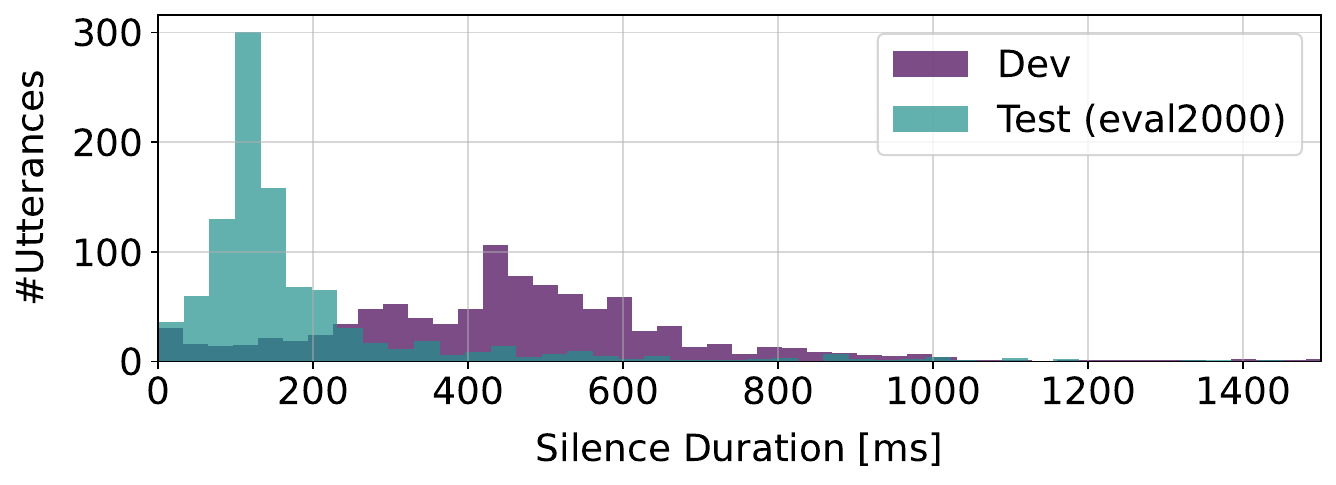}
    \vspace{-0.2cm}
    \caption{Distribution of silence duration $T - t_{\mathsf{EOU}}$ across development and test sets of Switchboard.}
    \vspace{-0.2cm}
    \label{fig:eos_eoa.png}
\end{figure}

\subsection{Task Formulation}
\label{ssec:2_problem_definition}
The objective of ASR is to predict an $N$-length token sequence $W \in \mathcal{V}^N$
from the corresponding $T$-length audio sequence $O \in \mathbb{R}^{T \times F}$,
where $\mathcal{V}$ is a vocabulary, and $F$ is the dimension of acoustic features in $O$.
Our ASR model aims to address the following predictive tasks,
where we detail the evaluation metrics used for each to enhance understanding.
Figure~\ref{fig:task} also illustrates a schematic drawing for each task.

\paragraph{Predictive EOU Detection} This task focuses on predicting the future endpoint of an utterance,
with its corresponding evaluation metric defined as $|\hat{t}_{\mathsf{EOU}} - t_{\mathsf{EOU}}|$,
where $\hat{t}_{\mathsf{EOU}}$ and $t_{\mathsf{EOU}}$ represent the predicted and actual times of the EOU, respectively.
The EOU is identified as the point in time when the user finished uttering the last word,
i.e., the waveform collapses~\cite{fan2023towards},
which is obtained by performing forced alignment.
It is important to differentiate this timing from the end of the audio file $T$,
as $t_\mathsf{EOU}$ generally occurs earlier than $T$ (i.e., $t_{\mathsf{EOU}} < T$).
In fact, in our experiments, we observe that the silence duration $T - t_\mathsf{EOU}$ can be up to \SI{1200}{\milli\second} in the datasets used.
Figure~\ref{fig:eos_eoa.png} depicts the distribution of this silence duration across the development and test sets of the Switchboard corpus.

\paragraph{Predictive ASR} This task aims to generate the full transcription before the utterance ends.
For evaluation,
we use the standard word error rate (WER) measured across the entire sentence, and
introduce a modified WER specifically for future predictions, which we refer to as FWER.
FWER is computed by using ground-truth tokens for the spoken content and calculating WER for the tokens predicted into the future.
This allows for the exclusive measurement of future predictions,
isolating them from the cumulative errors associated with the previously predicted tokens.

\begin{figure}
    \centering
    \vspace{-0.2cm}
    \includegraphics[width=0.9\linewidth]{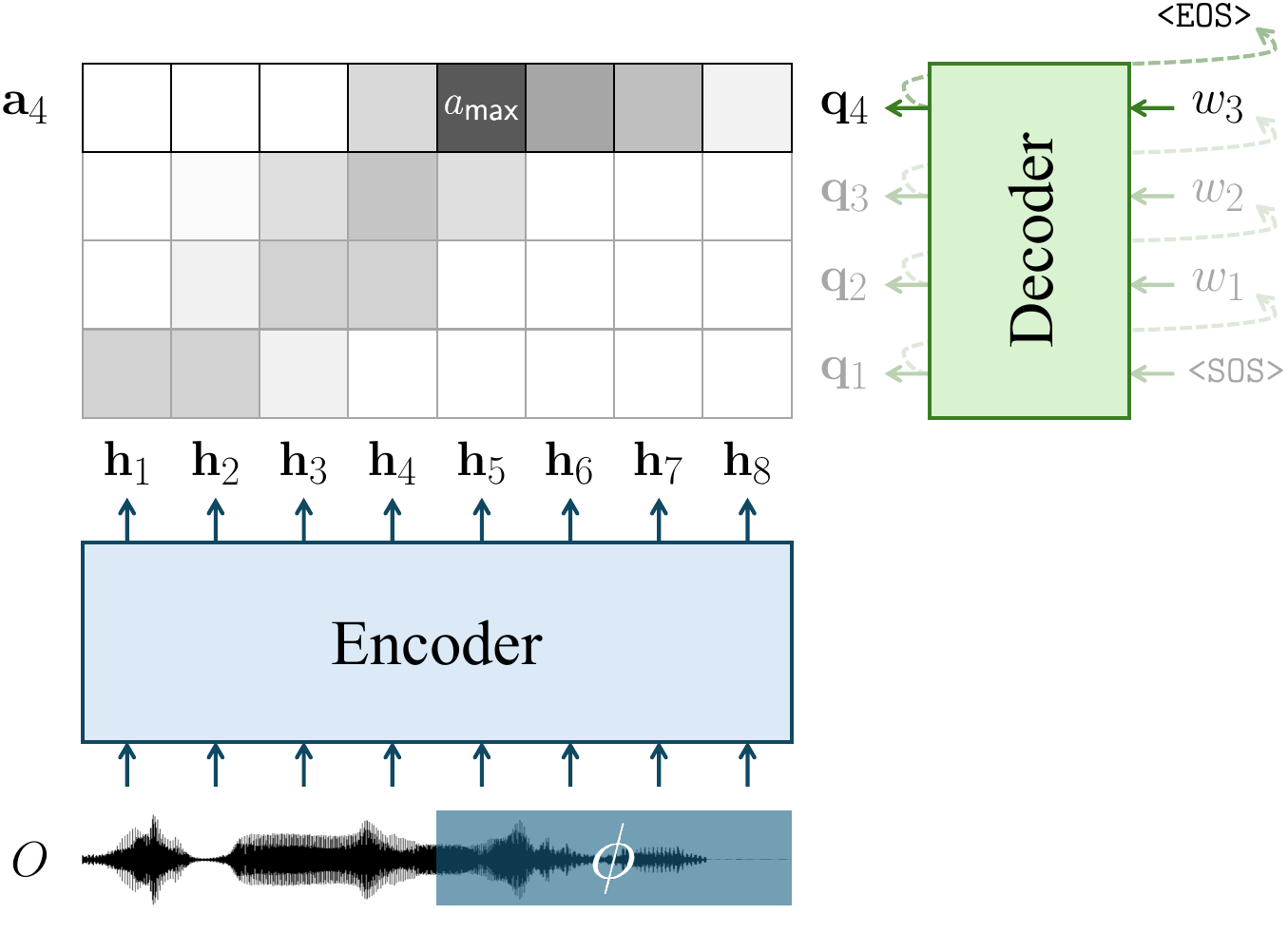}
    \vspace{-0.2cm}
    \caption{Proposed approach for detecting EOU based on cross-attention mechanism. It computes attention scores used for generating the final output, the end-of-sentence token ($\texttt{<EOS>}$). To identify the EOU, the upper boundary of the frames related to this final output is determined by comparing scores $a_t \in \mathbf{a}_4$ to the maximum score $a_{\mathsf{max}}$.}
    \label{fig:res}
\end{figure}

\subsection{Training Strategy based on Masked Future Input}
\label{ssec:mask_training}
To address the aforementioned predictive tasks,
we build an encoder-decoder-based ASR model~\cite{chorowski2015attention,chan2016listen,watanabe2017hybrid} that is explicitly trained to give the decoder an incentive to increase its acoustically conditional language model probabilities.
Specifically, we mask a segment of a training audio sample for a duration of $T_\mathsf{mask}$ milliseconds prior to the EOU time $t_\mathsf{EOU}$ and the trailing silence until the end of audio $T$,
as illustrated in Figure~\ref{fig:task}.
This training is expected to force the decoder to predict future tokens based on incomplete or absent acoustic information.

The proposed training approach is implemented by first extracting acoustic features $O$ from an audio sample.
We then randomly sample a masking duration $T_\mathsf{mask}$ from a uniform distribution spanning from $0$ to $M$.
In the final step, the acoustic features within the interval from $t_\mathsf{EOU} - T_\mathsf{mask}$ to $T$ are substituted with zero-vectors.
Importantly, we maintain positional encoding on the masked segment to assist the decoder in estimating the placement of future tokens relative to the unmasked part.

To address variations in the duration of silence after $t_\mathsf{EOU}$,
as indicated by Figure~\ref{fig:eos_eoa.png},
we also introduce variability in the audio input length.
This is achieved by sampling a duration from a uniform distribution between $-T_{\Delta}$ and $T_{\Delta}$ and
accordingly adjusting the length of the masked input by adding or removing zero vectors.

\subsection{Predictive EOU Detection Using Cross-Attention Weights}
\label{ssec:predictive_eou}
After training, our model performs EOU detection using the cross-attention mechanism,
as depicted in Figure~\ref{fig:res}.
Given an audio input $O$,
the encoder generates audio representations $H \in \mathbb{R}^{T \times D_{\mathsf{model}}}$,
and, subsequently,
the decoder produces token representations $Q \in \mathbb{R}^{(N + 1) \times D_{\mathsf{model}}}$.
Here, $D_{\mathsf{model}}$ denotes the dimensionality of the hidden layers within each network.
Notably, the ($N+1$)-th output from the decoder is specifically dedicated to predicting the end-of-sentence token.
During the computation of $Q$,
the model computes an attention score matrix $\mathbf{A} \in [0, 1]^{(N+1) \times T}$ by applying scaled dot-product attention against $H$~\cite{vaswani2017attention}.
Given the attention scores $\mathbf{a}_{N + 1} \in \mathbf{A}$ related to emitting the end-of-sentence token,
the maximum score in $\mathbf{a}_{N + 1}$ is defined as $a_{\mathsf{max}} = \max(\mathbf{a}_{N + 1})$.
Finally, the EOU time is estimated based on $a_{\mathsf{max}}$ as
\begin{equation}
    \hat{t}_{\mathsf{EOU}} = \tau \cdot \max\{ t \mid a_t \geq \Psi \cdot a_{\text{max}} \},
\end{equation}
where $\tau$ represents the duration of each encoder frame (i.e., $\tau = \SI{40}{\milli\second}$), and
the hyperparameter $\Psi$ is introduced to threshold the upper limit of the frames that receive attention.
For example in Figure~\ref{fig:res} with $T=8$ and $N=3$,
the attention scores in $\mathbf{a}_4$ can extend across the encoder outputs.
Since our goal is to identify the EOU,
we seek the rightmost frame associated with $\mathbf{q}_4$ based on the maximum score $a_{\mathsf{max}}$,
where the estimated EOU $\hat{t}_{\mathsf{EOU}}$ is likely to be at frame $6$, $7$, or $8$,
depending on the value of $\Psi$.

The above algorithm can work whether or not future content in the audio input is available.
When the entire input is accessible, it performs straightforward EOU detection.
On the other hand, in cases where future input is absent,
it becomes the predictive EOU task, which is our primary focus.

\subsection{Predictive ASR Using Decoder}
\label{ssec:predictive_asr}
The proposed model performs predictive ASR by processing audio input from the middle of an utterance,
utilizing the standard decoding algorithms typical of encoder-decoder-based ASR models.
Consistent with the training approach, additional zero-vector frames are appended to the input,
allowing the model to continue its autoregressive token generation of future content.
The length of these additional frames can vary,
thanks to the random sampling technique used to determine the mask and silence duration during training.
However, for evaluation purposes, we fill the input until it reaches the total length of the audio $T$ for simplicity.

\section{Experimental Setting}
All of the experiments were conducted using the codes and recipes provided by the ESPnet~\cite{watanabe2018espnet} toolkit.

\paragraph{Data}
We used the LibriSpeech (LS)~\cite{panayotov2015librispeech} and Switchboard (SWBD)~\cite{godfrey1992switchboard} datasets.
LS consists of single-speaker utterances extracted from read English audiobooks, and
we used the 100-hour subset (LS-100) for model training.
The utterances in LS can be characterized by their clear endpoints,
making them well-suited for model evaluation under ideal conditions.
SWBD includes single-speaker utterances derived from two-sided telephone conversations.
SWBD presents a more challenging scenario due to the dialogic nature,
involving complex turn-taking between speakers with ambiguous endpoints.
We ran the Montreal forced aligner~\cite{mcauliffe2017montreal} on the above datasets to obtain all timing annotations,
which were primarily used to compute the FWER (see Section~\ref{ssec:2_problem_definition}) and obtain target EOU.

\paragraph{Evaluated Models}
We trained our baseline model using the hybrid connectionist temporal classification (CTC) and attention model~\cite{watanabe2017hybrid},
featuring the encoder-decoder-based structure with auxiliary CTC loss applied to the encoder output.
The proposed model adopted the same architecture as the baseline model,
but it was trained using the masked audio input, as described in Section~\ref{ssec:mask_training}.
For both LS-100 and SWBD,
we used the Conformer-based network architecture~\cite{gulati2020conformer},
as implemented and defined by the corresponding ESPnet recipe\footnote{\url{https://github.com/espnet/espnet/blob/master/egs2/librispeech_100/asr1}}.

\paragraph{Training and Inference Configurations}
We adhered closely to the optimization configurations specified in the ESPnet recipe for each dataset.
For our training strategy based on masked future input (in Section~\ref{ssec:mask_training}),
we set $M = 500$ and $T_{\mathsf{\Delta}} = 200$ for both LS-100 and SWBD.
During EOU detection using cross-attention (in Section~\ref{ssec:predictive_eou}),
we set $\Psi = 0.1$ for LS and $\Psi = 1.0$,
based on the model’s performance in validation.
For ASR decoding (in Section~\ref{ssec:predictive_asr}),
we performed beam-search decoding with the beam size of $1$ or $20$.
To evaluate the predictive capability for both EOU detection and ASR,
we tested models using the mask durations of $T_{\mathsf{mask}} = 0$, $100$, $200$, $300$, $400$, and $500$.
Notably, for $T_{\mathsf{mask}} = 0$ all acoustic information is available and no predictive ASR is necessary.

\section{Results and Discussion}
\label{sec:experiments}
\begin{figure}
    \centering
    \includegraphics[width=\linewidth]{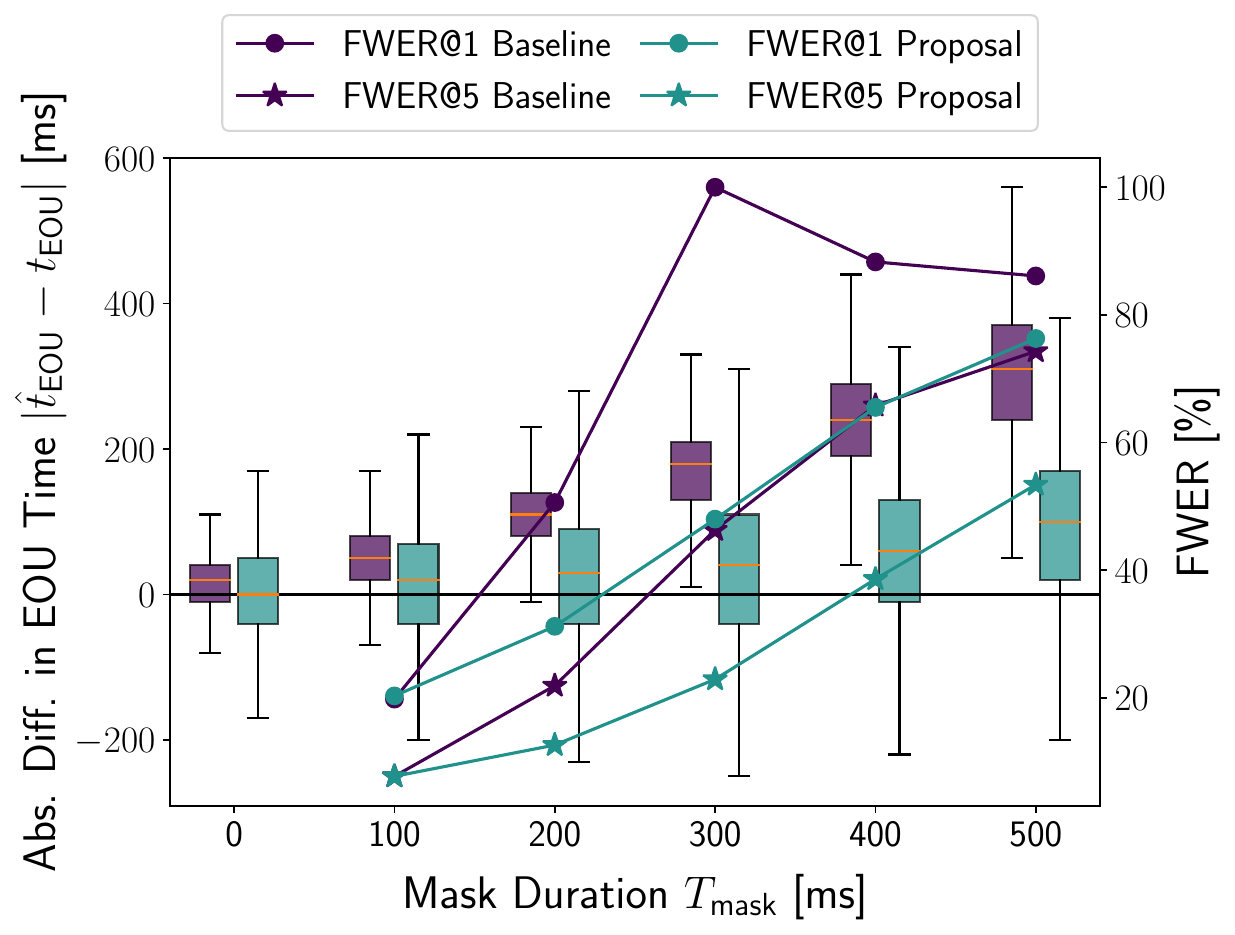}
    \vspace{-0.6cm}
    \caption{Absolute difference in EOU timing [ms] and FWER [\%] on LS-100 test set, evaluated across different mask durations.}
    \label{fig:koba_libri}
\end{figure}

\begin{figure}
    \centering
    \includegraphics[width=\linewidth]{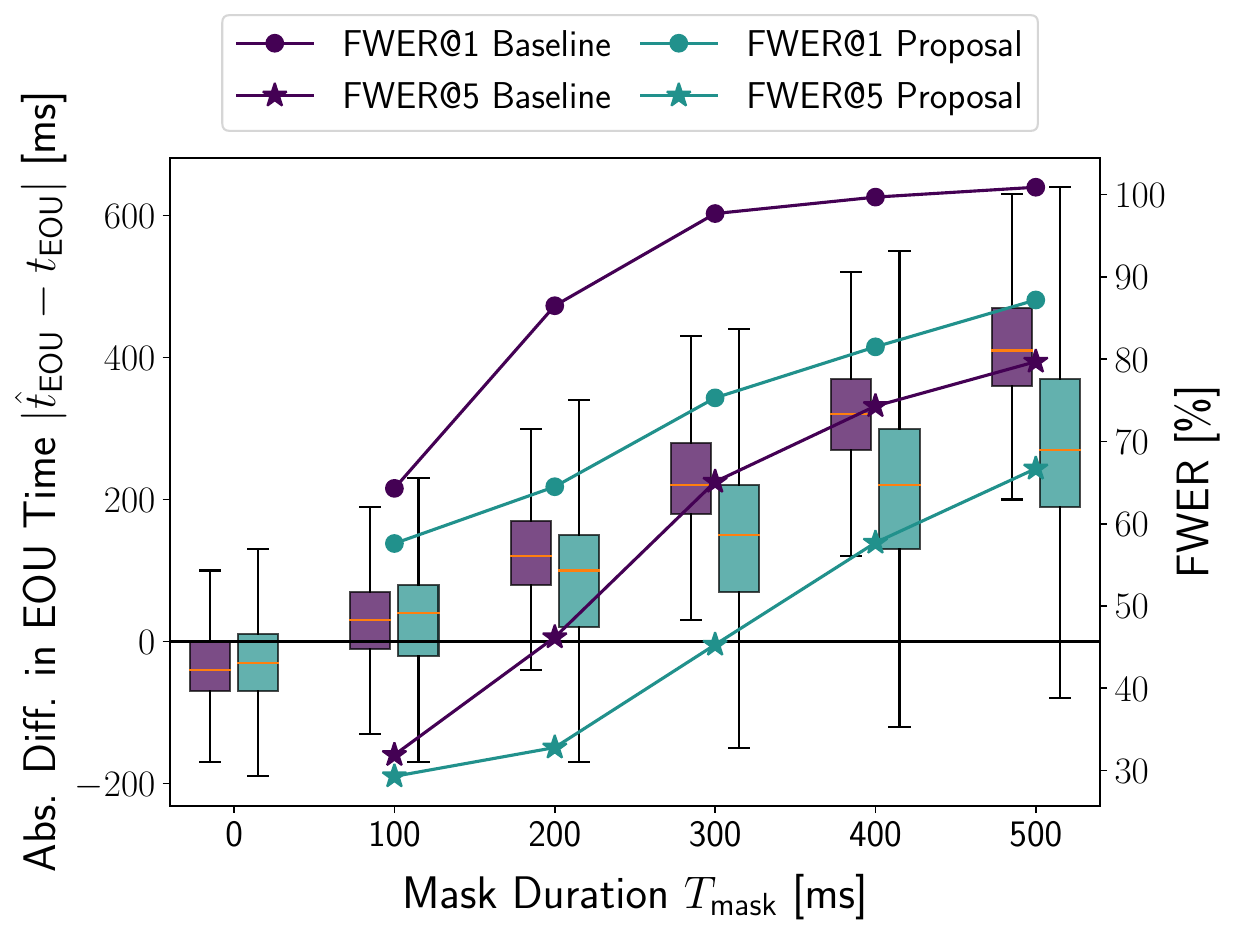}
    \vspace{-0.6cm}
    \caption{Absolute difference in EOU timing [ms] and FWER [\%] on SWBD test set, evaluated across different mask durations.}
    \vspace{-0.2cm}
    \label{fig:koba_swbd}
\end{figure}

\subsection{Predictive EOU Detection}
Figure~\ref{fig:koba_libri} reports box plots for LS-100,
showing the performance of our EOU detection using the cross-attention mechanism,
which was measured by the absolute difference between the predicted and ground-truth EOU timing $|\hat{t}_\mathsf{EOU} - t_\mathsf{EOU}|$ (as detailed in Section~\ref{ssec:2_problem_definition}).
When $T_\mathsf{mask} = 0$, indicating that the models had full access to the audio input,
EOU was detected reasonably well, with an average discrepancy of about \SI{0}{\milli\second}.
As the mask duration was increased, the error for the baseline model increased notably,
reaching an average difference of approximately \SI{300}{\milli\second} when $T_\mathsf{mask}=500$.
In contrast, the proposed model successfully mitigated this degradation especially for the higher mask durations,
suppressing the discrepancy around \SI{100}{\milli\second} at $T_\mathsf{mask}=500$.
This demonstrates the proposed training strategy,
which involves masking future input, was effective in enabling the model to perform the predictive EOU detection task.

Figure~\ref{fig:koba_swbd} shows results on SWBD, following a similar trend as observed in LS-100 between the baseline and proposed models.
However, the overall performance was inferior to LS-100, with greater variance in predictions.
This indicates increased challenges in forecasting the EOU in conversational speech,
where speaker terminations may be less distinct.
Nonetheless, the proposed model consistently outperformed the baseline,
particularly at the mask durations above \SI{200}{\milli\second}.

\subsection{Predictive ASR}
Figures~\ref{fig:koba_libri} and~\ref{fig:koba_swbd} plot the FWER results on LS-100 and SWBD, respectively,
which assessed WER solely based on future predictions.
We also present the FWER-at-5 (FWER@5) results,
obtained by performing beam search decoding to generate the top-five hypotheses and reporting the lowest FWER observed among these.
Notice that the error rates are generally high, and
this underscores the challenges of predicting upcoming tokens with various possible outcomes,
consistent with the findings reported in~\cite{yusuf2024speculative}.
Notably, the baseline model struggled with FWERs exceeding 80\% for mask durations longer than \SI{300}{\milli\second}.
In contrast, our model effectively reduced the errors thanks to the proposed training strategy,
which enhanced the decoder's capability to act as a generative language model, even in the absence of audio input.
By evaluating FWER@5, our model greatly improved performance, suppressing errors to below 70\%;
however, we note that this comes at the cost of requiring the NLP modules to handle responses for five potential ASR hypotheses.

\begin{table}[t]
    \centering
    \caption{WER [\%] on test sets for LS-100 and SWBD.}
    \vspace{-0.2cm}
    \label{fig:WER_table}
    \begin{tabular}{cccccccc}
    \toprule
    & & \multicolumn{6}{c}{\textbf{Mask Duration} $T_\mathsf{mask}$ [ms]} \\
    \cmidrule(l{0.3em}r{0.3em}){3-8}
    \textbf{Dataset} & \textbf{Model} & 0  & 100  & 200 & 300 & 400 & 500 \\
    \midrule
    \multirow{2}{*}[0pt]{LS-100} & Baseline &  8.5  &  8.8  &  10.5  &  12.3  &  13.4   &   14.3  \\
    & Proposal &  \textbf{8.4}  &  \textbf{8.6}  &  \textbf{9.2}  &  \textbf{10.3}  &  \textbf{11.7}   &  \textbf{13.2}  \\
    \midrule
    
    \multirow{2}{*}[0pt]{SWBD} & Baseline &  30.9  &   31.8   &  35.3   &   38.8  &  41.3   &   43.6  \\
    & Proposal &  \textbf{30.5}  &  \textbf{30.1}  &  \textbf{32.2}  &  \textbf{34.9}  &  \textbf{37.8}  &  \textbf{40.2}  \\
    \bottomrule
    \end{tabular}
\end{table}
Table~\ref{fig:WER_table} reports the WER for LS-100 and SWBD, computed for all words (not limited to future words) predicted by the models.
With the training strategy based on masking future input,
the proposed model consistently outperformed the baseline across various mask durations.
Interestingly, the proposed model exhibited superior performance when $T_{\mathsf{mask}}=0$.
This improvement can be attributed to enhancements in the decoder's ability to act as a language model,
which aided in learning dependencies among output tokens.

Overall, based on the findings presented, our model has successfully demonstrated its predictive capabilities.
We suggest that the model can reasonably operate up to \SI{300}{\milli\second} before an utterance ends,
providing extra time for the downstream NLP modules to prepare responses.

\section{Conclusion}
\label{sec:conclusion}
This paper proposed an ASR model that simulates human anticipatory capabilities
through the design of predictive EOU detection and predictive ASR tasks.
We developed a novel training strategy that involves randomly m asking future segments of an utterance,
thereby enabling the decoder to predict forthcoming words.
Additionally, we proposed a cross-attention-based algorithm that leverages alignment information to accurately determine the timing of the EOU.
The experimental results showed that our model is capable of predicting upcoming words and estimating EOU timing up to \SI{300}{\milli\second} prior to the actual EOU.

\paragraph{Acknowledgment}
This work was supported by the Baden-W\"urttemberg-STIPENDIUM.

\bibliographystyle{IEEEbib}
\bibliography{refs}

\begin{thebibliography}{10}

\bibitem{graves2012sequence}
Alex Graves,
\newblock ``Sequence transduction with recurrent neural networks,''
\newblock in {\em Proc. ICML Representation Learning Workshop}, 2012.

\bibitem{he2019streaming}
Yanzhang He, Tara~N Sainath, Rohit Prabhavalkar, Ian McGraw, Raziel Alvarez, Ding Zhao, David Rybach, Anjuli Kannan, Yonghui Wu, Ruoming Pang, et~al.,
\newblock ``Streaming end-to-end speech recognition for mobile devices,''
\newblock in {\em Proc. ICASSP}. IEEE, 2019, pp. 6381--6385.

\bibitem{zhang2020transformer}
Qian Zhang, Han Lu, Hasim Sak, Anshuman Tripathi, Erik McDermott, Stephen Koo, and Shankar Kumar,
\newblock ``Transformer transducer: A streamable speech recognition model with transformer encoders and {RNN-T} loss,''
\newblock in {\em Proc. ICASSP}, 2020, pp. 7829--7833.

\bibitem{moritz2020streaming}
Niko Moritz, Takaaki Hori, and Jonathan Le,
\newblock ``Streaming automatic speech recognition with the transformer model,''
\newblock in {\em Proc. ICASSP}, 2020, pp. 6074--6078.

\bibitem{tsunoo2021streaming}
Emiru Tsunoo, Yosuke Kashiwagi, and Shinji Watanabe,
\newblock ``Streaming transformer {ASR} with blockwise synchronous beam search,''
\newblock in {\em Proc. SLT}, 2021, pp. 22--29.

\bibitem{zhao2023mask}
Huaibo Zhao, Yosuke Higuchi, Yusuke Kida, Tetsuji Ogawa, and Tetsunori Kobayashi,
\newblock ``Mask-{CTC}-based encoder pre-training for streaming end-to-end speech recognition,''
\newblock in {\em Proc. EUSIPCO}, 2023, pp. 56--60.

\bibitem{narayanan2021cascaded}
Arun Narayanan, Tara~N Sainath, Ruoming Pang, Jiahui Yu, Chung-Cheng Chiu, Rohit Prabhavalkar, Ehsan Variani, and Trevor Strohman,
\newblock ``Cascaded encoders for unifying streaming and non-streaming {ASR},''
\newblock in {\em Proc. ICASSP}, 2021, pp. 5629--5633.

\bibitem{li2021better}
Bo~Li, Anmol Gulati, Jiahui Yu, Tara~N Sainath, Chung-Cheng Chiu, Arun Narayanan, Shuo-Yiin Chang, Ruoming Pang, Yanzhang He, James Qin, et~al.,
\newblock ``A better and faster end-to-end model for streaming {ASR},''
\newblock in {\em Proc. ICASSP}, 2021, pp. 5634--5638.

\bibitem{moritz2021dual}
Niko Moritz, Takaaki Hori, and Jonathan Le~Roux,
\newblock ``Dual causal/non-causal self-attention for streaming end-to-end speech recognition,''
\newblock in {\em Proc. Interspeech}, 2017, pp. 1909--1913.

\bibitem{zhao2023conversation}
Huaibo Zhao, Shinya Fujie, Tetsuji Ogawa, Jin Sakuma, Yusuke Kida, and Tetsunori Kobayashi,
\newblock ``Conversation-oriented {ASR} with multi-look-ahead {CBS} architecture,''
\newblock in {\em Proc. ICASSP}, 2023, pp. 1--5.

\bibitem{yu2021dual}
Jiahui Yu, Wei Han, Anmol Gulati, Chung-Cheng Chiu, Bo~Li, Tara~N Sainath, Yonghui Wu, and Ruoming Pang,
\newblock ``Dual-mode {ASR}: Unify and improve streaming {ASR} with full-context modeling,''
\newblock in {\em Proc. ICLR}, 2021.

\bibitem{strimel2023lookahead}
Grant Strimel, Yi~Xie, Brian~John King, Martin Radfar, Ariya Rastrow, and Athanasios Mouchtaris,
\newblock ``Lookahead when it matters: Adaptive non-causal transformers for streaming neural transducers,''
\newblock in {\em Proc. ICML}, 2023, pp. 32654--32676.

\bibitem{levinson2015timing}
Stephen~C Levinson and Francisco Torreira,
\newblock ``Timing in turn-taking and its implications for processing models of language,''
\newblock {\em Frontiers in psychology}, vol. 6, pp. 731, 2015.

\bibitem{ekstedt2020turngpt}
Erik Ekstedt and Gabriel Skantze,
\newblock ``{TurnGPT}: a transformer-based language model for predicting turn-taking in spoken dialog,''
\newblock in {\em Findings of EMNLP}, 2020, pp. 2981--2990.

\bibitem{sakuma2023response}
Jin Sakuma, Shinya Fujie, and Tetsunori Kobayashi,
\newblock ``Response timing estimation for spoken dialog systems based on syntactic completeness prediction,''
\newblock in {\em Proc. SLT}, 2023, pp. 369--374.

\bibitem{yusuf2024speculative}
Bolaji Yusuf, Murali~Karthick Baskar, Andrew Rosenberg, and Bhuvana Ramabhadran,
\newblock ``Speculative speech recognition by audio-prefixed low-rank adaptation of language models,''
\newblock in {\em Proc. Interspeech}, 2024, pp. 792--796.

\bibitem{chorowski2015attention}
Jan~K Chorowski, Dzmitry Bahdanau, Dmitriy Serdyuk, Kyunghyun Cho, and Yoshua Bengio,
\newblock ``Attention-based models for speech recognition,''
\newblock in {\em Proc. NeurIPS}, 2015, pp. 577--585.

\bibitem{chan2016listen}
William Chan, Navdeep Jaitly, Quoc Le, and Oriol Vinyals,
\newblock ``Listen, attend and spell: {A} neural network for large vocabulary conversational speech recognition,''
\newblock in {\em Proc. ICASSP}, 2016, pp. 4960--4964.

\bibitem{watanabe2017hybrid}
Shinji Watanabe, Takaaki Hori, Suyoun Kim, John~R Hershey, and Tomoki Hayashi,
\newblock ``Hybrid ctc/attention architecture for end-to-end speech recognition,''
\newblock {\em IEEE Journal of Selected Topics in Signal Processing}, vol. 11, no. 8, pp. 1240--1253, 2017.

\bibitem{chang2020low}
Shuo-Yiin Chang, Bo~Li, David Rybach, Yanzhang He, Wei Li, Tara~N Sainath, and Trevor Strohman,
\newblock ``Low latency speech recognition using end-to-end prefetching,''
\newblock in {\em Proc. Interspeech}, 2020, pp. 1962--1966.

\bibitem{fan2023towards}
Yifeng Fan, Colin Vaz, Di~He, Jahn Heymann, Viet~Anh Trinh, Zhe Zhang, and Venkatesh Ravichandran,
\newblock ``Towards accurate and real-time end-of-speech estimation,''
\newblock in {\em Proc. ICASSP}, 2023, pp. 1--5.

\bibitem{vaswani2017attention}
Ashish Vaswani, Noam Shazeer, Niki Parmar, Jakob Uszkoreit, Llion Jones, Aidan~N Gomez, {\L}ukasz Kaiser, and Illia Polosukhin,
\newblock ``Attention is all you need,''
\newblock in {\em Proc. NeurIPS}, 2017, pp. 5998--6008.

\bibitem{watanabe2018espnet}
Shinji Watanabe, Takaaki Hori, Shigeki Karita, Tomoki Hayashi, Jiro Nishitoba, Yuya Unno, Nelson Enrique~Yalta Soplin, Jahn Heymann, Matthew Wiesner, Nanxin Chen, et~al.,
\newblock ``Espnet: End-to-end speech processing toolkit,''
\newblock {\em arXiv preprint arXiv:1804.00015}, 2018.

\bibitem{panayotov2015librispeech}
Vassil Panayotov, Guoguo Chen, Daniel Povey, and Sanjeev Khudanpur,
\newblock ``Librispeech: An {ASR} corpus based on public domain audio books,''
\newblock in {\em Proc. ICASSP}, 2015, pp. 5206--5210.

\bibitem{godfrey1992switchboard}
John~J Godfrey, Edward~C Holliman, and Jane McDaniel,
\newblock ``{SWITCHBOARD}: Telephone speech corpus for research and development,''
\newblock in {\em Proc. ICASSP}, 1992, pp. 517--520.

\bibitem{mcauliffe2017montreal}
Michael McAuliffe, Michaela Socolof, Sarah Mihuc, Michael Wagner, and Morgan Sonderegger,
\newblock ``Montreal forced aligner: Trainable text-speech alignment using {Kaldi},''
\newblock in {\em Proc. Interspeech}, 2017, vol. 2017, pp. 498--502.

\bibitem{gulati2020conformer}
Anmol Gulati, James Qin, Chung-Cheng Chiu, Niki Parmar, Yu~Zhang, Jiahui Yu, Wei Han, Shibo Wang, Zhengdong Zhang, Yonghui Wu, and Ruoming Pang,
\newblock ``Conformer: Convolution-augmented transformer for speech recognition,''
\newblock in {\em Proc. Interspeech}, 2020, pp. 5036--5040.

\end{thebibliography}

\end{document}